\documentclass[twocolumn,showpacs,preprintnumbers,amsmath,amssymb,prb]{revtex4}
\usepackage{graphicx}
\begin{document}

\title{Low-temperature properties of the dilute dipolar magnet LiHo$_{x}$Y$_{1-x}$F$_4$ }
\author{A. Biltmo}
\author{P. Henelius}
\affiliation{Dept. of Theoretical Physics, Royal Institute of Technology, SE-106 91 Stockholm, Sweden}

\begin{abstract}
We analyze recent experiments on the dilute rare-earth compound LiHo$_{x}$Y$_{1-x}$F$_4$ in the context of an effective Ising dipolar model. Using a Monte Carlo method we calculate the low-temperature behavior of the specific heat and linear susceptibility, and compare our results to measurements. In our model the susceptibility follows a Curie-Weiss law at high temperature, $\chi\sim 1/(T-T_{\mbox{cw}})$, with a Curie-Weiss temperature that scales with dilution, $T_{\mbox{cw}}\sim x$, consistent with early experiments. We also find that the peak in the specific heat scales linearly with dilution, $C_{max}(T)\sim x$, in disagreement with recent experiments. Experimental studies do not reach a consensus on the functional form of these quantities, and in particular we do not see reported scalings of the form $\chi\sim T^{-0.75}$ and $\chi\sim \exp{(-T/T_0)}$. Furthermore we calculate the ground state magnetization as a function of dilution, and re-examine the phase diagram around the critical dilution $x_c=0.24\pm 0.03$. We find that the spin glass susceptibility for the Ising model does not diverge below $x_c$, while recent experiments give strong evidence for a stable spin-glass phase in LiHo$_{0.167}$Y$_{0.833}$F$_4$.
\end{abstract}

\date{\today}

\maketitle

\section{Introduction}

The rare-earth compound LiHoF$_4$ is used as a model magnet to investigate diverse magnetic phenomena such as quantum phase transitions,\cite{rose96} spin-glass behavior\cite{rose90} and quantum annealing\cite{broo99}. The magnetic behavior arises from the Ho$^{3+}$ ions which have tightly bound $4f$ electrons. This causes the exchange interaction to be weak, and the inter-ion interactions are predominantly dipolar. The local crystal field causes a strong anisotropy, and the interaction is Ising-like. To a first approximation LiHoF$_4$ is therefore believed to be good realization of a dipolar Ising model,\cite{chak04}
\begin{equation}
H=\frac{J}{2}\sum_{i\ne j}\frac{r_{ij}^2-3z_{ij}^2}{r_{ij}^5}
\sigma_i^z\sigma_j^z+ \frac{J_{\rm{ex}}}{2}\sum_{i,nn}\sigma^{z}_{i}\sigma^{z}_{
nn},
 \label{dipole}
\end{equation}
where we have used a dipolar coupling constant $J=0.214$ K and a nearest-neighbor ($nn$) exchange coupling $J_{\mbox{ex}}=0.12$ K.\cite{bilt07} The interspin distance is $r_{ij}$, with a component $z_{ij}$ along the Ising axis. The magnetic Ho$^{3+}$ ions sit on a tetragonal lattice with four ions per unit cell. To study quantum criticality a transverse magnetic field can be applied, and in order to study the effects of disorder the magnetic Ho$^{3+}$ ions can be substituted by nonmagnetic Y$^{3+}$ ions, resulting in LiHo$_{x}$Y$_{1-x}$F$_4$. During the last three decades LiHoF$_4$ has been extensively studied and used as a textbook example of a quantum magnet\cite{chak96, sach99}. However, in the case of substantial dilution measurements have reported a variety of functional forms for basic thermodynamic quantities, such as the static susceptibility and the specific heat.

The earliest data we find for the static susceptibility report a high-temperature Curie-Weiss scaling $\chi\sim 1/(T-T_{\mbox{cw}}) $ with Curie-Weiss temperatures $T_{\mbox{cw}}=0.05$ and 0.16 for dilution $x=0.045$ and 0.167 respectively.\cite{rose90} In a later work the susceptibility is found to diverge with a different power law, $\chi\sim T^{-0.75}$ ($x=0.045$),\cite{ghos03} and in a recent study the exponential low-temperature form $\chi=\exp(-T/T_0)$ is reported.\cite{jons07} The specific heat has also been measured by several different groups, and in an earlier study\cite{rose90} of the specific heat a peak was found at about $T=0.3 $ K for $x=0.045$, while there was only a much broader maximum below $T=0.2$ K for x=0.167.\cite{rose90} A later study by the same group finds peaks at $T=0.1$ K, as well as at $T=0.3$ K, for $x=0.045$.\cite{ghos03} A more recent study\cite{quil06} displays a dilution independent maximum in the specific heat at about $T=0.1$ K for $x= 0.018$, 0.045 and 0.08. 
 
Finally, the nature of the glassy phase at low temperatures has also been the topic of several experimental studies. Earlier work found a spin-liquid (anti-glass) phase at extreme dilution (x=0.045), followed by a stable spin-glass phase at dilution x=0.167, and finally a magnetic phase at x=0.3.\cite{rose90} More recent experiments did not detect a spin glass transition,\cite{jons07} but this may have been due to the use of large magnetic fields.\cite{rose08} Recent numerical work on dilute dipoles on a small cubic lattice fails to find a spin glass transition,\cite{snid05} and so does a recent numerical study of the above model for LiHo$_{x}$Y$_{1-x}$F$_4$.\cite{bilt07} 

In this study we confine ourselves to the case of no external magnetic field, but it is interesting to note that quantum Monte Carlo studies of the above non-diluted model including an applied transverse field\cite{chak04, tabe08_2} do not reach good agreement with the experimental phase diagram, even for small transverse fields.

Much of the recent theoretical work on LiHo$_{x}$Y$_{1-x}$F$_4$ has focused on the effects of the hyperfine coupling and off-diagonal dipolar terms resulting in corrections to the above Hamiltonian.\cite{sche05,tabe06,tabe08,sche08} Yet a non-perturabative calculation beyond mean-field of several fundamental properties such as the specific heat and linear susceptibility is lacking even for the first-order model described by Eq.~(\ref{dipole}). The goal of the present work is to numerically investigate the above model and determine to what extent it can be used to interpret the experimental results. In particular we calculate the static susceptibility and specific heat and compare our result to recent experiments. We also calculate the ground state magnetization as a function of temperature in order to get an independent estimate of the critical dilution, $x_c$, where the magnetization vanishes. Finally, we reexamine the low-temperature disordered phase and search for evidence of a stable spin-glass phase.

\section{Method}

We have used a single spin flip Monte Carlo method and applied periodic boundary conditions. To handle the long-range nature of the interaction we have used the Ewald summation method\cite{Ewald:1921} as explained in an earlier study.\cite{bilt07} To overcome energy barriers in the glassy phase and reach lower temperatures than in previous work we have used the replica exchange Monte Carlo method.\cite{swen86} The method involves simulating an ensemble of systems at suitably chosen temperatures $T_i$, and the algorithm has two main phases. In the first phase each replica is independently evolved in (Monte Carlo) time using the single spin Metropolis algorithm. In the second phase attempts are made to exchange the replicas at adjacent temperatures $T_i$ and $T_{i+1}$. A full Monte Carlo step consists of one attempted spin flip per spin (on average) followed by ten attempts to exchange neighboring replicas. For the simulation to converge at low temperatures it is important that the swap rate of the replicas is not too low. Theory and empirical studies\cite{EarDee2005, KonKof2005} have shown the optimal rate to be around $20 \%$ and these simulations were carried out with swap rates $\gtrsim 20 \%$ .


In the present study we have calculated the specific heat,
\begin{equation}
C=\frac{1}{k_B T^2}\left( \langle H^2\rangle -\langle H \rangle^2\right),
\end{equation}
 and the magnetic susceptibility
\begin{equation}
\chi=\frac{1}{T}\left( \langle M^2\rangle -\langle M\rangle^2\rangle\right),
\end{equation}
where the magnetization $M=\sum_{i=1}^{N}\sigma^z_i$.

In order to study the disordered phase we have calculated the Edwards-Anderson overlap between two replicas,\cite{bind86}
$$q=\frac{1}{N} \sum_i \sigma_{i,1}^{z}\sigma_{i,2}^{z} $$ and the corresponding Binder ratio
$$g_q=1-\frac{ \langle q^4\rangle }{3\langle q^2\rangle^2}.$$
The spin glass susceptibility is defined as $ \chi_{SG}= \langle q^2\rangle /T^2$.
In addition to the thermal average we have calculated an average of 400-600 quenched disorder realizations.

\section{Results}
\begin{figure}[htp]
\resizebox{\hsize}{!}{\includegraphics[type=eps, ext=.eps, read=.eps,clip=true]{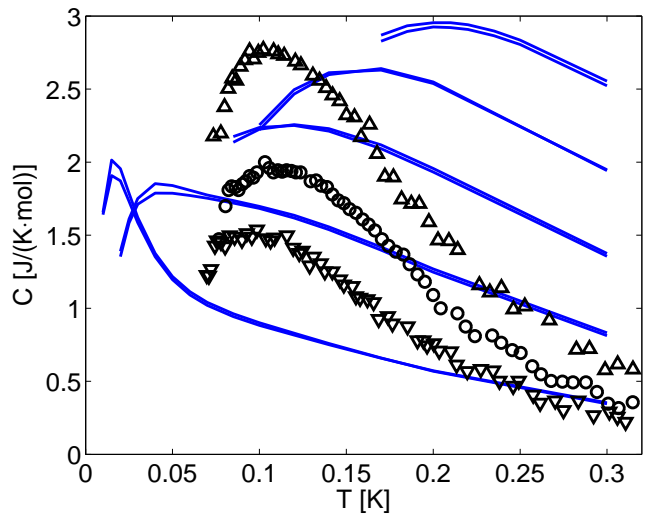}}
\caption{Specific heat as a function of temperature. Experimental data from Ref.~\onlinecite{quil06} (open symbols) for dilution x=0.08, x=0.045 and x=0.018 (top to bottom). Monte Carlo results (solid lines) for dilution x=0.165, 0.12, 0.08, 0.045 and 0.018 (top to bottom). To display the limited finite-size effects results for $8^3$ and $10^3$ (upper curve) unit cells are shown for each dilution.}
\label{specheat}
\end{figure}

\begin{figure}[htp]
\resizebox{\hsize}{!}{\includegraphics[type=eps, ext=.eps, read=.eps,clip=true]{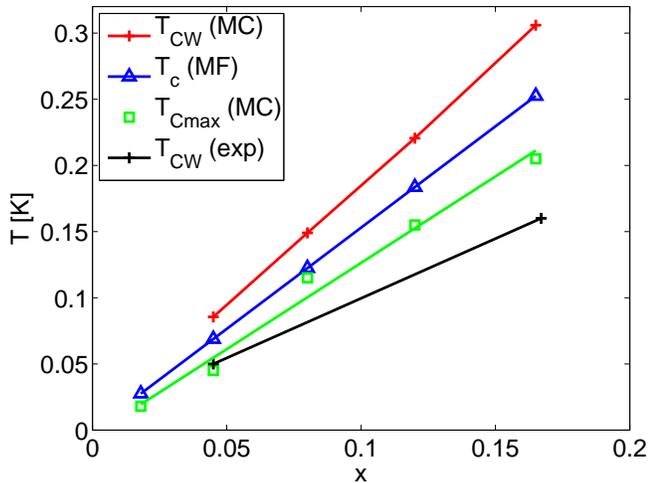}}
\caption{ Curie-Weiss temperature (Monte Carlo), mean-field critical temperature, maximum in specific heat (Monte Carlo), and experimental Curie-Weiss temperature from Ref.~\onlinecite{rose90} (top to bottom) as a function of dilution.}
\label{max} 
\end{figure}

First we will compare our calculation of the specific heat to experimental data. In Fig.~\ref{specheat} we show our results for the specific heat as a function of temperature. In the same figure we also show recent experimental data.\cite{quil06} There is qualitative agreement and both sets of curves indicate that the specific heat grows with decreasing dilution. Both sets of curves exhibit a maximum for some intermediate temperature, but the experimental peak position is roughly independent of the dilution, while the calculated peak position scales linearly with $x$, which can be seen if Fig.~\ref{max}. In an Ising spin glass the specific heat exhibits a broad maximum in the vicinity of the transition temperature,\cite{bind86} and as the mean-field transition temperature scales linearly with dilution, we might expect the scaling observed in our calculation. However, the present experimental data are for high dilution, and experimentally there is no spin glass transition in the limit of extreme dilution. It would therefore be of interest to measure the specific heat for less dilute systems, to see whether the expected linear increase in the peak position of the susceptibility is recovered in this limit. Furthermore, the total specific heat is dominated by the contribution from the nuclear spins\cite{quil06} and the necessary subtraction required to obtain the experimental data is sensitive to the form of the subtracted single-ion contribution. Any uncertainty in the noninteracting specific heat could cause a big change in the resulting electronic specific heat.

\begin{figure}[htp]
\resizebox{\hsize}{!}{\includegraphics[type=eps, ext=.eps, read=.eps,clip=true]{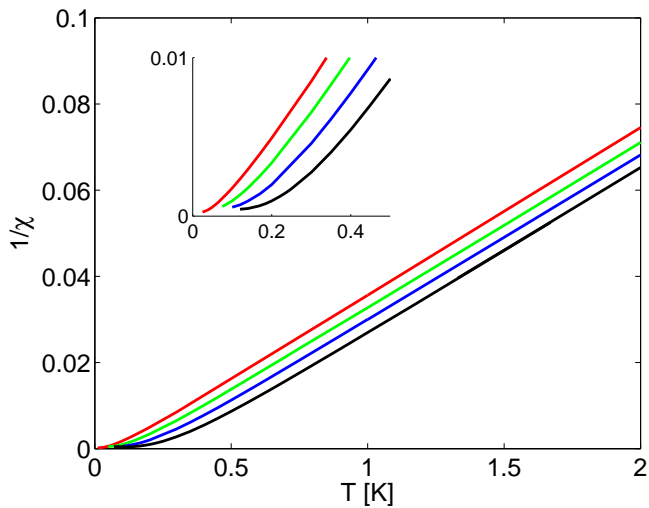}}
\caption{ Inverse susceptibility per spin as a function of temperature for x=0.045, 0.08, 0.12 and 0.167 (top to bottom). The displayed data has converged in system size.}
\label{invsusc} 
\end{figure}
\begin{figure}[htp]
\resizebox{\hsize}{!}{\includegraphics[type=eps, ext=.eps, read=.eps,clip=true]{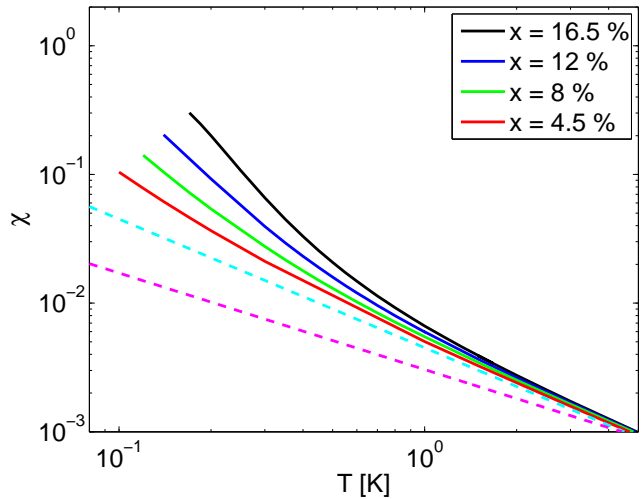}}
\caption{Susceptibility per spin as a function of temperature for x=0.167, 0.12, 0.08 and 0.045 (top to bottom, solid lines). The dashed lines have slope -1 and -0.75 (lower line).}
\label{logsusc} 
\end{figure}

Next we will analyze our results for the linear susceptibility. The inverse susceptibility is plotted in Fig.~\ref{invsusc}. In good qualitative agreement with early measurements of the susceptibility\cite{rose90} we see Curie-Weiss behavior, $\chi\sim 1/(T-T_{cw})$, at higher temperatures, and deviations at low temperatures. As the dilution is increased the susceptibility approaches the free spin limit $\chi\sim T^{-1}$ as expected. Extrapolating the Curie scaling to the intercept gives us the Curie-Weiss temperature, $T_{\mbox{cw}}$, which is positive, in accordance with the ferromagnetic correlations in LiHoF$_4$. As can be seen in Fig.~\ref{max} we find that $T_{\mbox{cw}}$ scales linearly in $x$, and we get $T_{\mbox{cw}}=0.08$ and 0.30 for x=0.045 and 0.167 respectively. Experiments reported in Ref.~\onlinecite{rose90} 
found that $T_{\mbox{cw}}=0.05$ and 0.16 for x=0.045 and 0.167 respectively. From Fig.~\ref{max} we see that while our calculated Curie-Weiss temperature is higher than the mean-field critical temperature, the experimental results are lower. We therefore reach qualitative, but not quantitative, agreement with this set of experiments. However, there is no experimental consensus on the functional form of the susceptibility and a later set of measurements by the same group report a scaling of the form $\chi\sim T^{-0.75}$ for x=0.045.\cite{ghos03} In order to further analyze the functional form we plot our results for the susceptibility in a log-log plot in Fig.~\ref{logsusc}. From the inserted straight lines we see the Curie scaling $\chi\sim T^{\alpha}$ with $\alpha=-1$ at higher temperatures. As the temperature is lowered the susceptibility diverges faster, with an exponent $\alpha < -1$, contrary to reported measurements $\alpha=-0.75$.\cite{ghos03} The experimental data was explained by off-diagonal terms in the dipolar interaction that arise when the material is diluted.
Our omission of these terms could explain the discrepancy, but the fact remains that our results agree quite well with the earlier measurements of the susceptibility. Furthermore, the reported scaling of $\chi\sim T^{-0.75}$ persists up to $T=2$ K, and given that the average diagonal local dipolar field is of the order $1.53\times 0.045 \approx 0.07$ K it is surprising that there are deviations from Curie scaling at such elevated temperatures. Finally, experimental data for the susceptibility has also been argued to be well modeled by an exponential, low-temperature form\cite{jons07} $\chi=\exp(-T/T_0)$. Plotting our results for $ \chi $ in a semilog plot does not result in a straight line over any significant temperature interval. 

\begin{figure}[htp]
\resizebox{\hsize}{!}{\includegraphics[type=eps, ext=.eps, read=.eps,clip=true]{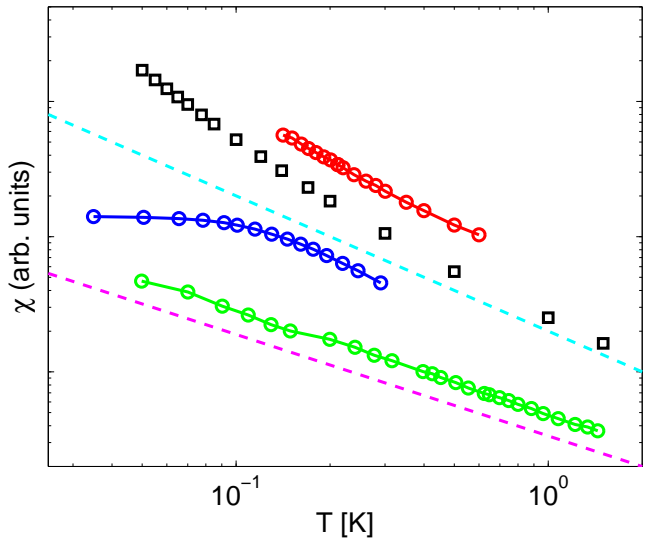}}
\caption{Susceptibility at x=0.045. From top to bottom the open symbols are experimental data from Ref.~\onlinecite{rose90}, Monte Carlo results, experimental data from Ref.~\onlinecite{jons07} and Ref.~\onlinecite{ghos03}. The dashed lines have slope -1 and -0.75 (lower curve). The curves have been separated vertically. }
\label{susc045}
\end{figure}

In order to compare the various results for the static susceptibility for the extreme dilution $x=0.045$ we display all the measurements in Fig.~\ref{susc045}. We have shifted the curves vertically to display the functional form better. We see that data from Ref.~\onlinecite{ghos03} follows the form $\chi\sim T^{-0.75}$ over the whole temperature range from 0.05 K to 2K. In the high temperature limit our calculation, as well as data from Ref.~\onlinecite{rose90} and Ref.~\onlinecite{jons07} tend to the Curie scaling $\chi\sim T^{-1}$. At low temperature our calculation and data from Ref.~\onlinecite{rose90} diverge faster than $T^{-1}$ while data from Ref.~\onlinecite{jons07} grows significantly more slowly. Due to this discrepancy between different experiments it is difficult to determine how well the classical dipolar Ising model reflects the magnetic behavior of LiHo$_x$Y$_{1-x}$F$_4$ in the high-dilution limit. More measurements that could explain the above experimental differences would be necessary in order to draw more definite conclusions.

\begin{figure}[htp]
\resizebox{\hsize}{!}{\includegraphics[type=eps, ext=.eps, read=.eps,clip=true]{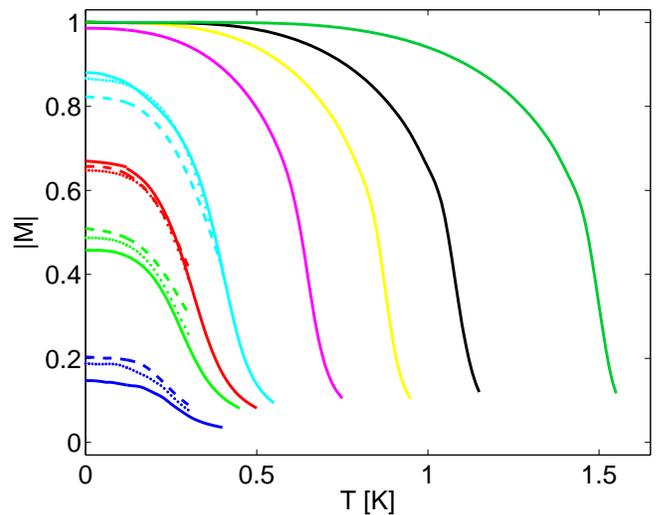}}
\caption{ Magnetization per spin as a function of temperature for dilution x=0.25, 0.3, 0.33, 0.375, 0.5, 0.625, 0.75 and 1 (left to right). At low dilution the result is for $10^3$ unit cells, while for x=0.3, 0.33 and 0.375 the system sizes are $6^3$ (dashed line), $8^3$ (dotted line) and $10^3$ unit cells. For the highest dilution (x=0.25) the system sizes are $12^3$ (dashed line), $14^3$ (dotted line) and $16^3$ unit cells. }
\label{mag}
\end{figure}

\begin{figure}[htp]
\resizebox{\hsize}{!}{\includegraphics[type=eps, ext=.eps, read=.eps,clip=true]{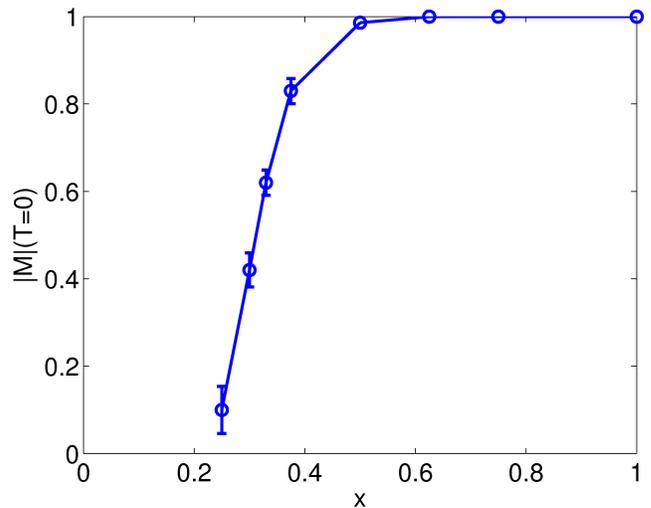}}
\caption{Ground state magnetization per spin as a function of dilution.}
\label{gsmag}
\end{figure}

\begin{figure}[htp]
\resizebox{\hsize}{!}{\includegraphics[type=eps, ext=.eps, read=.eps,clip=true]{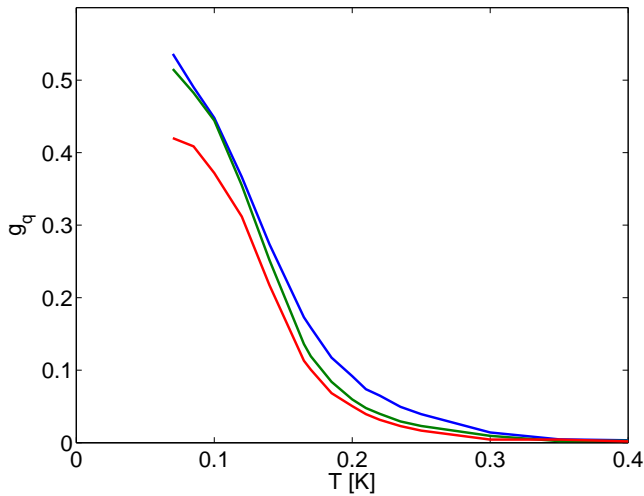}}
\caption{Binder ratio for the spin overlap at dilution x=0.167. System sizes are $10^3$, $12^3$ and $14^3$ unit cells (top to bottom).}
\label{Binder}
\end{figure}

\begin{figure}[htp]
\resizebox{\hsize}{!}{\includegraphics[type=eps, ext=.eps, read=.eps,clip=true]{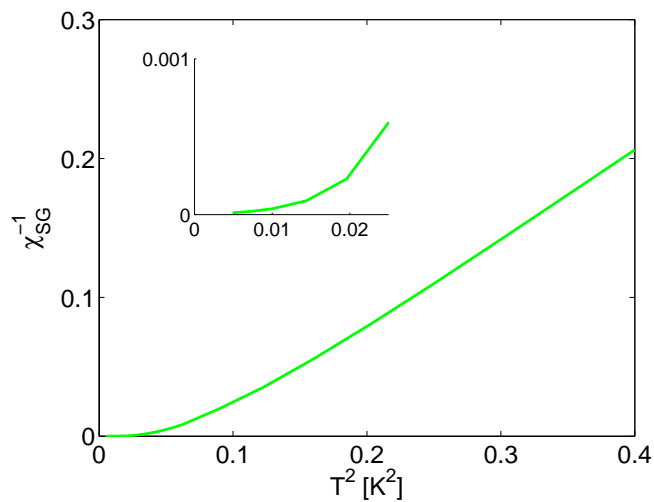}}
\caption{Inverse spin glass susceptibility at dilution $x=0.167$. The displayed data has converged in system size.}
\label{sgsusc}
\end{figure}

Next we consider the magnetization of the dilute model. Using a parallel tempering method we are able to determine the magnetization curves to lower temperature than in a previous study,\cite{bilt07} as can be seen in Fig.~\ref{mag}. Notice that, for a given dilution, the magnetization increases with system size for low dilution ($x=0.375$), while it decreases in the more dilute systems ($x=0.25$ and $x=0.30$). Extrapolating to the ground state we obtain the ground state magnetization curve in Fig.~\ref{gsmag}. Finite size effects and statistical errors prevent us from a very exact determination, but the curve indicates that the the critical concentration $x_c$, where the magnetization vanishes, is about $x_c=0.24\pm 0.03$, which is a bit higher than the value $x_c=0.21\pm0.02$ reported in a previous calculation\cite{bilt07}. We are not aware of any precise experimental determination of $x_c$, but our result is consistent with experiments that report a spin-glass phase at $x=0.167$, but a ferromagnetic state at $x=0.3$.\cite{rose90} Our results also compare well with a previous zero-temperature Monte Carlo study of Ising dipoles on a diluted BCC lattice,\cite{xu91} where it was found that $x_c=0.3\pm 0.1$. 

Finally we consider the disordered phase for $x < x_c$. We calculate the Binder ratio for the spin overlap, $g_q$, in the disordered phase. If there is a stable glass phase the curves for different system sizes are expected to cross at the freezing temperature. In a previous study\cite{bilt07} no crossing was found, indicating that there is no freezing of the spin glass. Here we have repeated the calculation using the parallel tempering method in order to obtain more reliable data in the highly disordered phase. The result is shown in Fig.~\ref{Binder}, and as can be seen we still detect no crossing for $x=0.167$. In order to analyze the nature of the disordered phase further we also consider the spin glass susceptibility $\chi_{SG}$. In a study of the Heisenberg spin glass it was argued that the Binder ratio of the spin overlap may not intersect at the freezing temperature for all boundary conditions.\cite{star98} The study suggests that the divergence of the spin glass susceptibility may be a better indicator of the freezing transition. Since many properties of the long-range dipolar model are quite sensitive to the choice of boundary conditions, we therefore show results for the inverse spin glass susceptibility in Fig.~\ref{sgsusc}. The finite size effects are very small and the spin glass susceptibility does not diverge at a finite temperature. Since there is quite convincing experimental evidence for a spin-glass transition\cite{rose08} at $x=0.167$ the results are puzzling and either there are some aspects of the simulations of the glassy dipolar phase that differ from the short-range Ising spin glass, or the neglected off-diagonal terms in the Hamiltonian are necessary to stabilize the glassy phase observed in LiHo$_{0.167}$Y$_{0.833}$F$_4$.

\section{Discussion and conclusion}

In this work we have made direct comparisons between calculations done on a first-order effective Ising dipolar model and experimental data obtained for LiHo$_{x}$Y$_{1-x}$F$_4$. We have focused on the static susceptibility and specific heat. Probably due to the slow dynamics in the highly disordered phase, different sets of experiments do not agree very well. Nevertheless, our calculation agrees well with static susceptibility measurements of Ref.~\onlinecite{rose90} in the highly disordered regime. Obtaining non-perturbative results beyond mean-field for the first-order classical dipolar model is an essential step on the way to understanding the physical properties of LiHo$_{x}$Y$_{1-x}$F$_4$. Three particularly puzzling experimental results that cannot be explained by our calculation on the classical dipolar model are the susceptibility scaling $\chi\sim T^{-0.75}$ for $x=0.045$, a specific heat maximum that is independent of dilution, and the finite temperature spin-glass transition. 

The difference between our calculation and the experimental results may be explained by quantum mechanical terms that are not included in our classical model. We have ignored the hyperfine coupling between nuclear and electronic spins. In the low-temperature limit this is generally important, particularly in the presence of a external transverse field. In this study we do not consider an applied magnetic field, and the hyperfine coupling is only expected to re-normalize the inter-spin coupling.\cite{sche05} Neglecting the hyperfine coupling should therefore not lead to quantitatively new results, but may explain some of the quantitative differences between our calculations and the experiments. We have also ignored off-diagonal terms in the dipolar Hamiltonian, which result in an effective random transverse field.\cite{sche05, tabe06,sile07} Including these terms in the calculation would be an important next step in interpreting the measurements on LiHo$_{x}$Y$_{1-x}$F$_4$.

\section{Acknowledgements}

We thank M. Gingras and S. Girvin for useful discussions. This work was supported by the G\"oran Gustafsson foundation and the Swedish Research Council.


\bibliography{artikel2}

\end{document}